\documentclass[aps,prl,preprint,superscriptaddress]{revtex4}

\usepackage{amsbsy} 

\usepackage{epsfig} 

\begin{document}

\title{Fermi surfaces of iron-pnictide high-$T_c$ superconductors from the limit of local magnetic moments}


\author{J.P. Rodriguez}

\affiliation{Department of Physics and Astronomy, 
California State University, Los Angeles, California 90032}

\author{M.A.N. Araujo}

\affiliation{Departamento de F\'{\i}sica, Universidade de \'Evora, P-7000-671, \'Evora,
Portugal}

\affiliation{CFIF, Instituto Superior
T\'ecnico, TU Lisbon, Av. Rovisco Pais, 1049-001 Lisboa, Portugal}
\date{\today}

\author{P.D. Sacramento}

\affiliation{CFIF, Instituto Superior
T\'ecnico, TU Lisbon, Av. Rovisco Pais, 1049-001 Lisboa, Portugal}
\date{\today}

\begin{abstract}
A 2-orbital $t$-$J$ model over the square lattice
that describes low-energy electronic excitations 
in iron-pnictide high-$T_c$ superconductors 
is analyzed with Schwinger-boson-slave-fermion
mean field theory and by exact numerical diagonalization on a finite
 system.
When inter-orbital hole hopping is suppressed,
a quantum critical point (QCP) is identified that separates
a commensurate spin-density wave (cSDW) state
at strong Hund's rule coupling from a hidden half-metal state at weak Hund's rule coupling.
Low-energy spinwaves that disperse anisotropically from cSDW momenta are predicted at the QCP.
Nested Fermi surfaces similar to those observed experimentally in iron-pnictide materials
are also predicted in such case.
\end{abstract}

\maketitle

\section{I. Introduction}
High-temperature superconductivity in iron-pnictide materials is achieved by 
injecting charge carriers into
stoichiometric parent compounds
that show commensurate spin-density wave (cSDW) order 
over a tetragonal lattice of iron atoms\cite{new_sc}\cite{delacruz}.
The charge carriers in these systems 
have predominantly iron $3d$-orbital character\cite{haule_08}\cite{ma_08}\cite{graser_09}, 
and they exist at two-dimensional (2D) hole-type 
and electron-type Fermi surface pockets centered, respectively, at 
zero 2D momentum and at the 2D cSDW momenta $(h/2a){\bf\hat x}$(${\bf\hat y}$)
\cite{fink09}\cite{brouet09}\cite{brouet11}.
Here, $a$ is the iron 2D lattice constant.
Density functional theory (DFT) calculations correctly account for the Fermi surfaces
that are also  observed  in the cSDW\cite{dong_08},
but they predict an ordered moment of approximately 2 Bohr magnetons ($\mu_B$) 
that is large compared to measured values that can be as low as $0.3\, \mu_B$ \cite{delacruz}.
Frustrated 2D Heisenberg models that assume local magnetic moments at the iron atoms
can account for the weak cSDW observed in parent compounds, 
on the other hand\cite{Si&A}\cite{jpr_ehr_09}\cite{oitmaa_10}\cite{thalmeier_10}\cite{jpr10}. 
They cannot predict the observed Fermi surface pockets, however,
because local-moment magnets are Mott insulators.

Below, we introduce mobile holes into a two-orbital local-moment model
for frustrated magnetism
that successfully describes the weak cSDW nature
of iron-pnictide systems\cite{jpr10}.
Exact diagonalization of one hole that hops over a square lattice of iron atoms in a single layer
obtains a robust cSDW groundstate at 
$1/2$ the cSDW momenta
when Hund's rule is obeyed (cf. ref. \cite{kane89}).
Both the exact and a mean-field analysis find that two Fermi surface hole pockets centered
at zero 2D momentum can emerge
when Hund's rule is violated,
 at sufficiently weak inter-orbital hopping of electrons.
The ground state in question is the half-metal state
with antiferromagnetic order across
the $3d_{(x+iy)z}$ and $3d_{(x-iy)z}$
orbitals of the iron atom\cite{jpr10}.
Proximity to a quantum critical point (QCP) that separates
the hidden half-metal at weak Hund's coupling 
from a cSDW metal\cite{ran_09} at strong Hund's coupling
results in weak cSDW order and in  low-energy spin-wave excitations at cSDW wave numbers.
The latter disperse in a manner
that is consistent with
inelastic neutron scattering (INS) measurements
in iron-pnictide metals\cite{zhao_09}\cite{hayden_10}.
The critical spinwaves in turn  result in
nested Fermi surface pockets
that are centered at cSDW wave numbers,
in agreement with angle-resolved photoemission (ARPES)
on iron-pnictide metals\cite{brouet09}.
These have mixed electron-type and hole-type character, however,
which is a feature that has also been seen in ARPES studies
of iron-pnictide metals\cite{brouet11}.


\section{II. 2-Orbital $t$-$J$ model: mean field theory and exact results}
We shall now show that a hidden half-metal groundstate emerges from
the following two-orbital $t$-$J$ model over the square lattice,
where double occupancy at a site-orbital is strictly forbidden:
\begin{equation}
H = -\sum_{\langle i,j \rangle} \sum_{\alpha , \beta} \sum_{s} (t_1^{\alpha,\beta} \tilde c_{i, \alpha,s}^{\dagger} \tilde c_{j,\beta,s} + {\rm h.c.})
+{1\over 2} J_0 \sum_i \biggl[\sum_{\alpha} {\bf S}_{i, \alpha}\biggr]^2 +
    \sum_{\langle i,j \rangle} \sum_{\alpha , \beta} J_1^{\alpha,\beta}
                                      {\bf S}_{i, \alpha} \cdot {\bf S}_{j, \beta} +
    \sum_{\langle\langle i,j \rangle\rangle} \sum_{\alpha , \beta} J_2^{\alpha,\beta}
                                      {\bf S}_{i, \alpha} \cdot {\bf S}_{j, \beta}.
\label{tj0j1j2}
\end{equation}
Above, ${\bf S}_{i,\alpha}$ is the spin operator that acts on the spin $s_0 = 1/2$ state of the
$3d_{(x+iy)z}$ ($d+$) or $3d_{(x-iy)z}$ ($d-$) orbital, 
$\alpha = 0$ or $1$,
in the iron atom at site $i$.  The latter runs over the square lattice
of iron atoms that make up an isolated layer.  
The former basis of $d\pm$ orbitals
is the least localized one
 (cf. ref. \cite{e-r_63}),
which maximizes the Hund's coupling, $-J_0$.
It therefore minimizes the Hund's rule exchange energy 
in the 2-orbital $t$-$J$ model (\ref{tj0j1j2}).
Nearest neighbor and next-nearest neighbor Heisenberg exchange
across the links $\langle i,j\rangle$ and $\langle\langle i,j\rangle\rangle$
is controlled by the 
exchange coupling constants
$J_1^{\alpha,\beta}$ and $J_2^{\alpha,\beta}$, respectively.  
These are necessarily isotropic over the $d\pm$ orbital basis.
Correlated hopping of an electron in orbital $\alpha$ to a neighboring unoccupied 
orbital $\beta$ is controlled by the 
hopping matrix element $t_1^{\alpha,\beta}$.
The Heisenberg model that corresponds to (\ref{tj0j1j2}) 
in the absence of charge carriers 
possesses a QCP at large $s_0$ 
that separates a cSDW at strong Hund's coupling
from a hidden ferromagnet 
that shows $\nwarrow_{d+}\searrow_{d-}$ spin order
at weak Hund's coupling 
if off-diagonal frustration exists\cite{jpr10}: e.g.
$J_1^{\parallel} = 0$, $J_1^{\perp} > 0$, and  $J_2^{\parallel} = J_2^{\perp} > 0$.
Here the superscripts $\parallel$ and $\perp$ 
refer to the relationship between the orbital indices $\alpha$ and $\beta$.  
Recent DFT calculations show that the superposition of direct ferromagnetic
exchange with super-exchange across nearest-neighbor iron atoms
can result
in the cancellation  $J_1^{\parallel} = 0$ \cite{ma_08}.
The remaining positive exchange coupling constants are assumed to be due to
the super-exchange mechanism\cite{Si&A}\cite{anderson_50}.

Let us now turn off inter-orbital hopping: $t_1^{\perp} = 0$.  
Notice that 
antiferromagnetic order across the $d+$ and $d-$ orbitals
then remains intact in the presence of mobile holes\cite{jpr10}.
This is a classical 
picture for a half-metal groundstate that
shows $\nwarrow_{d+}\searrow_{d-}$ spin order.
To describe it,
we adopt the Schwinger-boson ($b$) slave-fermion ($f$) representation 
for the correlated electron\cite{kane89}\cite{Auerbach_Larson_91}:
$\tilde c_{i,\alpha,s} = b_{i,\alpha,s} f_{i,\alpha}^{\dagger}$
and 
${\bf S}_{i,\alpha} =
(1/2) \sum_{s,s{\prime}}
f_{i,\alpha} b_{i,\alpha,s}^{\dagger} 
{\boldsymbol\sigma}_{s,s^{\prime}} b_{i,\alpha,s^\prime} f_{i,\alpha}^{\dagger}$,
with the constraint
%
\begin{equation}
2 s_0 = b_{i,\alpha,\uparrow}^{\dagger} b_{i,\alpha,\uparrow} 
+ b_{i,\alpha,\downarrow}^{\dagger} b_{i,\alpha,\downarrow} 
+ f_{i,\alpha}^{\dagger} f_{i,\alpha}
\label{no_double_occupy}
\end{equation}
imposed at each site-orbital to exclude double occupancy.  
From here on we set $\hbar = 1$.
Following Arovas and Auerbach\cite{Arovas_Auerbach_88}, 
we next rotate about the spin $y$ axis by an angle $\pi$ 
on one of the antiferromagnetic sublattices
in order to decouple the spins:
$b_{i,0,\uparrow} \rightarrow - b_{i,0,\downarrow}$ and
$b_{i,0,\downarrow} \rightarrow b_{i,0,\uparrow}$.
Again following these authors,
we now define mean fields that are set by the pattern 
of ferromagnetic ($\parallel$) versus antiferromagnetic ($\perp$) pairs of neighboring spins:
$Q_0^{\perp} = \langle b_{i,\alpha,s} b_{i,\beta,s}\rangle$,
$Q_{1(2)}^{\parallel} = \langle b_{i,\alpha,s}^{\dagger} b_{j,\alpha,s}\rangle$ and
$Q_{1(2)}^{\perp} = \langle b_{i,\alpha,s} b_{j,\beta,s}\rangle$
for (next) nearest-neighbor links, where $\alpha\neq\beta$.
We add to that list the mean field 
$P_1^{\parallel} = {1\over 2} \langle f_{i,\alpha}^{\dagger} f_{j,\alpha}\rangle$
for nearest-neighbor hopping of holes within the same orbital.  
The mean-field approximation for the $t$-$J$ model Hamiltonian (\ref{tj0j1j2}) then has the form
$H_{MF} = H_0[Q,P] + H_b + H_f$, 
where $H_0[Q,P]$ consolidates the bilinear terms among the mean fields,
 where 
%
$$H_b = {1\over 2}\sum_{k}\sum_{s} \{ 
\Omega_{\parallel}(k) [b_s^{\dagger}(k) b_s(k) +  b_s(-k) b_s^{\dagger}(-k)]
            + \Omega_{\perp}(k) [b_s^{\dagger}(k) b_s^{\dagger}(-k) + b_s(-k) b_s(k)]\}$$
is the Hamiltonian for free Schwinger bosons, with
\begin{eqnarray*}
\Omega_{\parallel}(k) &=& 
\delta\lambda + 
\sum_{n=0,1,2} z_n J_n^{\prime\perp} Q_n^{\perp}
- 4 (J_1^{\prime\parallel} Q_1^{\parallel} + 2 t_1^{\parallel} P_1^{\parallel})[1-\gamma_1({\bf k})]
- 4 J_2^{\prime\parallel} Q_2^{\parallel}[1-\gamma_2({\bf k})]\\
\Omega_{\perp}(k) &=&
-e^{i k_0} \sum_{n=0,1,2} z_n J_n^{\prime\perp} Q_n^{\perp}\gamma_n({\bf k}) ,
\end{eqnarray*}
and where
$H_f = \sum_k \varepsilon_f(k) f^{\dagger}(k) f(k)$ 
is the Hamiltonian for free slave fermions,
with 
$\varepsilon_f(k) = 8t_1^{\parallel} Q_1^{\parallel} \gamma_1({\bf k}) - \mu$.
Above, 
$k = (k_0,{\bf k})$ 
is the  3-momentum for these excitations,
with corresponding destruction operators
$b_s(k) = {\cal N}^{-1/2}\sum_{\alpha = 0}^1\sum_i e^{i(k_0 \alpha + {\bf k}\cdot{\bf r}_i)} b_{i,\alpha,s}$ and
$f(k) = {\cal N}^{-1/2}\sum_{\alpha = 0}^1\sum_i e^{i(k_0 \alpha + {\bf k}\cdot{\bf r}_i)} f_{i,\alpha}$.
Here $k_0 = 0, \pi$ represent even and odd superpositions of the
$d\pm$ orbitals,
while ${\cal N} = 2 N_{\rm Fe}$ denotes the number of sites-orbitals 
on the square lattice of iron atoms.
Also above, 
$z_0 = 1$ and $z_{1(2)} = 4$,
$\gamma_0({\bf k}) = 1$ and
$\gamma_{1(2)}({\bf k}) = {1\over 2} ({\rm cos}\, k_{x(+)} a\,+\,{\rm cos}\, k_{y(-)} a)$, where
$k_{\pm} = k_x \pm k_y$, while
$\delta\lambda$ is the boson chemical potential 
that enforces the constraint against double occupancy (\ref{no_double_occupy}) on {\it average}
 over the bulk of the system.  
The concentration of mobile  holes per site-orbital, $x$,
sets the chemical potential of the slave fermions, $\mu$.
Last, the effect of mobile holes on the Heisenberg
spin-exchange is accounted for by 
the effective exchange coupling  constants\cite{Auerbach_Larson_91}
$J^{\prime} = (1-x)^2 J$. 

The solution to the above mean field theory is achieved by making 
the standard Bogoliubov transformation 
of the boson field,
$b_s(k) = ({\rm cosh}\, \theta_k) \beta_s(k) + ({\rm sinh}\, \theta_k) \beta_s^{\dagger}(-k)$,
 with
${\rm cosh} \, 2\theta = \Omega_{\parallel} / \omega_b$ and
${\rm sinh} \, 2\theta = -\Omega_{\perp} / \omega_b$, where
$\omega_b = (\Omega_{\parallel}^2 - \Omega_{\perp}^2)^{1/2}$
is the energy eigenvalue of the ($\beta$) boson.  
Enforcing the constraint against double occuppancy (\ref{no_double_occupy})
on average then yields the principal mean field equation\cite{Arovas_Auerbach_88}
\begin{equation}
s_0 + {1\over 2} - {1\over 2}x = 
{\cal N}^{-1} \sum_k ({\rm cosh}\, 2 \theta_k) (n_B[\omega_b(k)]+{1\over 2}),
\label{principal}
\end{equation}
where $n_B$ denotes the Bose-Einstein distribution.  
Ideal Bose-Einstein condensation (BEC) into the
two lowest-energy states at ${\bf k} = 0$ occurs
as temperature
$T\rightarrow 0$, in which case $\delta\lambda\rightarrow 0$.  
The remaining self-consistent
equations for the Schwinger-boson mean fields are 
\begin{eqnarray*}
Q_n^{\parallel} &=& 
{\cal N}^{-1} \sum_k \gamma_n({\bf k}) ({\rm cosh}\, 2 \theta_k) (n_B[\omega_b(k)]+{1\over 2})\quad
{\rm and}\\
Q_n^{\perp} &=&
{\cal N}^{-1} \sum_k \gamma_n({\bf k}) e^{ik_0} ({\rm sinh}\, 2 \theta_k) (n_B[\omega_b(k)]+{1\over 2}).
\end{eqnarray*}
%
After comparison with (\ref{principal}),
ideal BEC 
as $T \rightarrow 0$ implies the
unique value $Q = s_0$ for all five of these mean fields 
in the large-$s_0$
limit.
Last, the self-consistent mean field equation for intra-orbital hole hopping is 
$P_1^{\parallel} = {\cal N}^{-1} \sum_{\bf k} \gamma_1({\bf k}) n_F[\varepsilon_f({\bf k})]$,
where $n_F$ is the Fermi distribution function.
We henceforth assume a hole band at low doping, $t_1^{\parallel} < 0$ and  $x \ll 1$,
which implies two degenerate circular Fermi surfaces centered at
zero 2D momentum with Fermi wave vector $k_F a = (4\pi x)^{1/2}$.
This yields the amplitude $P_1^{\parallel} = x/2$ for intra-orbital hole hopping.
Inspection of the spectrum for Schwinger bosons, $\omega_b({\bf k})$,
yields 
a spin gap 
at cSDW wave numbers
$(\pi/a) {\bf\hat x}$(${\bf\hat y}$)
equal to
\begin{equation}
\Delta_{cSDW} = (1-x)^2 (2 s_0) [(4 J_2^{\perp} - J_{0c}) (J_0 - J_{0c})]^{1/2},
\label{delta_csdw}
\end{equation}
where
$- J_{0c} = 2 (J_1^{\perp} - J_1^{\parallel}) - 4 J_2^{\parallel} -
  (1-x)^{-2} s_0^{-1} 2  t_1^{\parallel} x$
is the critical Hund's coupling at which $\Delta_{cSDW}\rightarrow 0$.
Notice that intra-orbital hole hopping
stabilizes the hidden half-metal state.
We therefore  propose
({\it i}) that the normal state of iron-pnictide superconductors 
is described by the present hidden half-metal state, and 
({\it ii}) that the  cSDW/superconductor transition 
that these systems commonly exhibit\cite{delacruz}\cite{brouet09}
is controlled by the QCP at Hund's coupling $-J_{0c}$.
The linear increase of $- J_{0c}$ with the concentration of holes $x$ 
implies a charge-carrier-poor cSDW and a charge-carrier-rich superconductor,
which is consistent with experiment.

\begin{figure}
\includegraphics[scale=0.70, angle=-90]{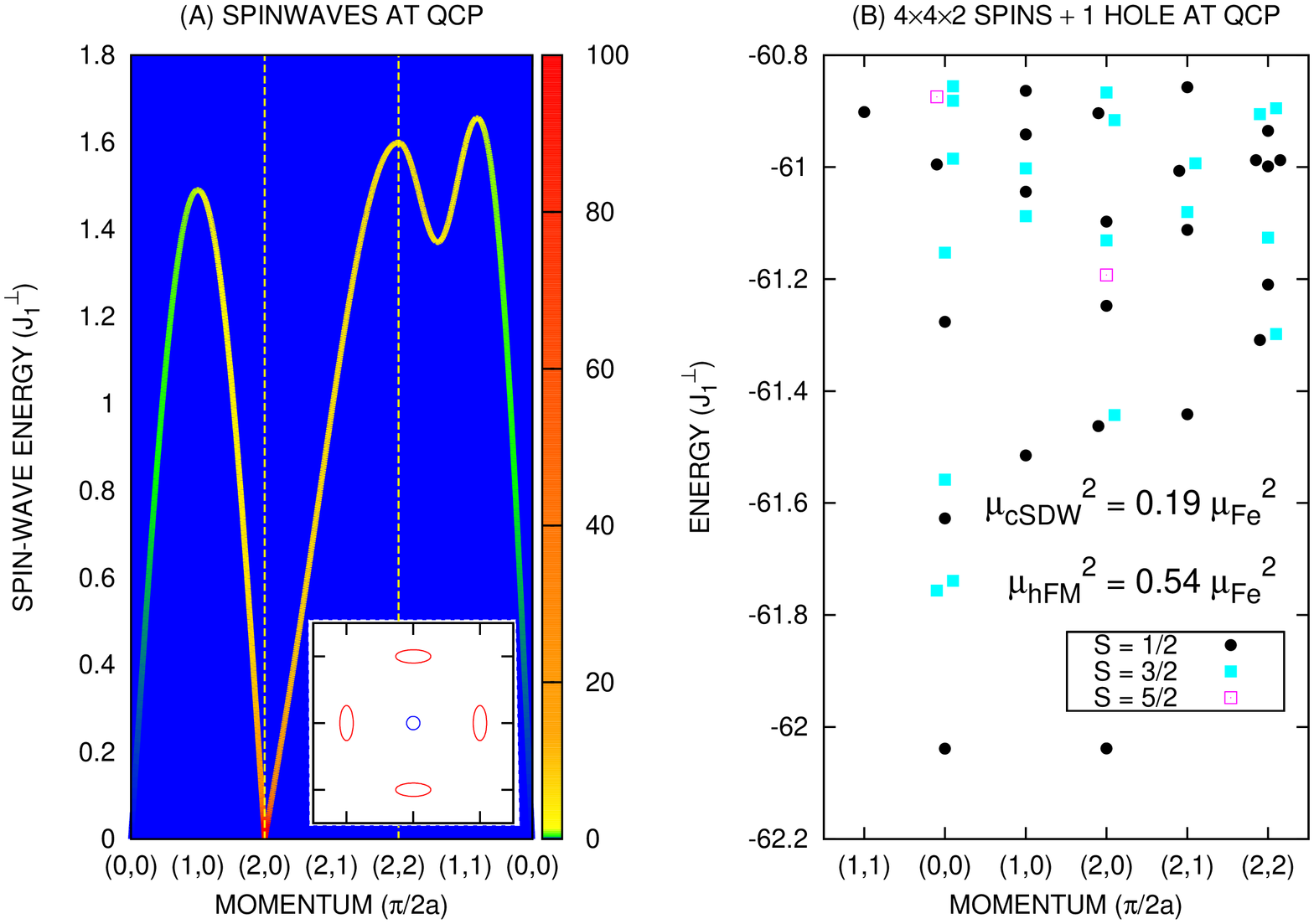}
\caption{The mean-field result for the dynamical spin response function
in the zero-temperature-large-$s_0$ limit, Eq. (\ref{chi_perp}), 
is evaluated (a) with the following set of parameters:
$J_1^{\parallel} = 0$, $J_1^{\perp} > 0$,
$J_2^{\parallel} = 0.3\, J_1^{\perp} = J_2^{\perp}$,
$t_1^{\parallel} = -5 J_1^{\perp}$, $t_1^{\perp} = 0$, $x = 0.01$, $s_0 = 1/2$, and
$J_0 = J_{0c}$.
Low-energy contours are displayed in the inset. 
Also shown (b) is the low-energy spectrum of the corresponding  $t$-$J$ model,
 Eq. (\ref{tj0j1j2}),  over a
$4\times 4\times 2$  lattice with one hole
at $J_0 = - 2.27\, J_1^{\perp}$.
Ordered magnetic moments
over the groundstate (at zero 2D momentum)
are also listed there:
$\langle \mbox{\boldmath$\mu$} (k)\cdot \mbox{\boldmath$\mu$} (-k)\rangle_0$,
where
$\mbox{\boldmath$\mu$} (k) = [2\mu_B/(N_{\rm Fe}-{1\over 2})]
 \sum_{\alpha=0}^1 \sum_i e^{i(k_0\alpha + {\bf k}\cdot{\bf r}_i)} {\bf S}_{i,\alpha}$,
and where 
$k = (\pi,0,0)$ and $(0,\pi/a,0)$, respectively,
for hidden ferromagnetic (hFM) and for cSDW order.
In general, $\mu_{\rm Fe}^2 = (33/31)(2\mu_B)^2$.}
\label{sw_spctrm}
\end{figure}

Transverse dynamical spin correlations are obtained directly from 
the above Schwinger-boson-slave-fermion mean field theory.  
In particular, we have
%
$$\langle S_{x(y)}  S_{x(y)}\rangle |_{k_0,{\bf k},\omega} =  
{1\over 2}(1-x)^2 [G_b*G_b^* +(-) F_b*F_b^*]|_{\pi(0)+k_0,{\bf k},\omega},$$
%
where 
$i G_b(k,\omega) = \langle b_s(k,\omega) b_s^{\dagger}(k,\omega)\rangle$ and
$i F_b(k,\omega) = \langle b_s(k,\omega) b_s (-k,-\omega)\rangle$ 
are the regular and the anomalous Greens functions for the Schwinger bosons, 
and where the notation  $f*g$ denotes a convolution in frequency and momentum.
This yields an Auerbach-Arovas expression for 
the dynamical spin correlator at $T>0$ \cite{Auerbach_Arovas_88}.
It is easily evaluated in the zero-temperature limit,
where ideal BEC
of the Schwinger bosons into the 
doubly degenerate ${\bf k} = 0$ ground state occurs.  
It contributes to half of the net transverse spin correlator,
which in this limit and at large $s_0$ reads 
\begin{equation}
i \langle {\bf S}_{\perp} \cdot {\bf S}_{\perp} \rangle |_{k,\omega}
= (1-x)^2 s_0 (\Omega_{+}/\Omega_{-})^{1/2} 
([\omega_b(k) - \omega ]^{-1} + [\omega_b(k) + \omega ]^{-1}).
\label{chi_perp}
\end{equation}
Here, $\Omega_{\pm} = \Omega_{\parallel}\pm\Omega_{\perp}$.
The above dynamical spin correlator coincides with the transverse spin
susceptibility, $\chi_{\perp}(k,\omega)$, in the present zero-temperature limit 
by the fluctuation-dissipation theorem.
Observe now that 
$\Omega_{-}(\pi,{\bf k})$ and $\Omega_{+}(0,{\bf k})$
both vanish at ${\bf k} = 0$ in general, 
while 
$\Omega_{-}(0,{\bf k})$ and $\Omega_{+}(\pi,{\bf k})$
both  vanish at cSDW wave vectors
${\bf k} = (\pi/a) {\bf\hat x}$(${\bf\hat y}$)
at the QCP \cite{jpr10}.
The identity $\omega_b = (\Omega_- \Omega_+)^{1/2}$
then ultimately  yields that spinwaves at zero 2D momentum disperse isotropically as 
$\omega_b({\bf k}) = v_0 |{\bf k}|$,
while those at cSDW wave numbers disperse anisotropically as 
\begin{equation}
\omega_b({\bf k}) =
[v_{0}^2 (k_{l} - \pi / a)^2 + v_{0}^2 (k_{t} / \gamma_{cSDW})^2 + \Delta_{cSDW}^2]^{1/2}.
\label{sdw_dispersion}
\end{equation}
(See fig. \ref{sw_spctrm}a.)
Here, $k_l$ and $k_t$ are the longitudinal and the transverse components of ${\bf k}$
with respect to the cSDW wave number.
The longitudinal spin-wave velocity is
\begin{equation}
v_0 = 2 s_0 a (1-x)^2 ([J_1^{\perp} - J_1^{\parallel}(x) + 2 (J_2^{\perp} - J_2^{\parallel})]
\cdot [{1\over 2} J_0 + 2 J_1^{\perp} + 2 J_2^{\perp}])^{1/2}
\label{v0}
\end{equation}
and the anisotropy parameter is
\begin{equation}
\gamma_{cSDW} = ([2(J_2^{\parallel} + J_2^{\perp}) + J_1^{\parallel}(x) + J_1^{\perp}]/
[2(J_2^{\parallel} + J_2^{\perp}) - J_1^{\parallel}(x) - J_1^{\perp}])^{1/2},
\label{ani_csdw}
\end{equation}
which is greater than unity.
Here  $J_1^{\parallel}(x) = J_1^{\parallel} +  (1-x)^{-2} s_0^{-1} t_1^{\parallel} x$.
Study of the spectral weights in  expression (\ref{chi_perp}) for $\chi_{\perp}(k,\omega)$ 
then yields that the former spinwaves at zero 2D momentum are hidden ($k_0 = \pi$),
while that the latter spinwaves at cSDW momenta are observable ($k_0 = 0$).
Figure \ref{sw_spctrm}a displays $\chi_{\perp}(k,\omega)$ at the QCP  in the observable channel, $k_0 = 0$,
assuming off-diagonal magnetic frustration and a low concentration of mobile holes: 
$J_1^{\parallel} = 0$, $J_1^{\perp} > 0$, 
$J_2^{\parallel} = 0.3 J_1^{\perp} = J_2^{\perp}$,
$t_1^{\parallel} = -5 J_1^{\perp}$,
$t_1^{\perp} = 0$ and $x = 0.01$.
Setting $s_0 J_1^{\perp} \sim 70$ meV
yields a successful fit\cite{jpr10} 
to spin-wave spectra
obtained from INS
on the superconductor BaFe$_{2-x}$Co$_x$As$_2$ \cite{hayden_10}. 
The critical spin-wave dispersion in fig. \ref{sw_spctrm}a also
notably shows a local maximum at the N\' eel momentum $(\pi/a,\pi/a)$,
which agrees with INS
on the parent compound CaFe$_2$As$_2$ \cite{zhao_09}.
Proximity to the QCP also naturally accounts for the low values of 
the magnetic moment associated with cSDW order ($\mu_{cSDW}$) 
that are seen in iron-pnictide parent compounds by neutron diffraction\cite{delacruz}.
Indeed, 
fig. \ref{sw_spctrm}b displays
exact results for one hole roaming over a $4\times4$ square lattice of iron atoms at the QCP,
where $\mu_{cSDW}$
is a fraction of the maximum possible ordered moment achieved in the true ferromagnetic state,
$\mu_{\rm Fe} = (33/31)^{1/2} 2 \mu_{B}$. (See end of section.) 
Last, Eq. (\ref{chi_perp}) coincides with the 
large-$s_0$ result 
for $\chi_{\perp}(k,\omega)$
obtained by one of the authors
in the hidden ferromagnetic state at $x = 0$ \cite{jpr10}.

The electronic structure of the hidden half-metal state 
can also be obtained directly from the above
Schwinger-boson-slave-fermion mean field theory.  
In  particular, the electron propagator
is given by the convolution of the propagator for 
Schwinger bosons with the propagator for slave fermions:
$i G(k,\omega) = G_b * G_f^* |_{k,\omega}$, 
where $i G_f(k,\omega) = \langle f(k,\omega) f^{\dagger}(k,\omega)\rangle$.
A standard summation of Matsubara frequencies  yields the expression 
\begin{equation}
G(k,\omega) = {1\over{\cal N}}\sum_q \Biggl[ ({\rm cosh}\, \theta_q)^2
{{n_B[\omega_b (q)] + n_F[\varepsilon_f (q-k)]}\over{\omega - \omega_b(q) + \varepsilon_f(q-k)}} +
({\rm sinh}\, \theta_q)^2
{{n_B[\omega_b (q)] + n_F[-\varepsilon_f(q-k)]}\over{\omega + \omega_b(q) + \varepsilon_f(q-k)}}\Biggr] .
\label{G}
\end{equation}
Ideal BEC
of the Schwinger bosons at 2D momentum ${\bf q} = 0$
results in the following
coherent contribution to the electronic spectral function
at zero temperature and at large $s_0$:
${\rm Im}\, G_{\rm coh}(k,\omega) = s_0 
{\pi} \delta[\omega + \varepsilon_f(k)]$.
It reveals the two degenerate hole bands expected from the classical picture
of a
half-metal state with $\nwarrow_{d+}\searrow_{d-}$ spin order
and with no inter-orbital hopping\cite{jpr10}.
The fermion contribution to ${\rm Im}\, G(k,\omega)$ above
represents incoherent excitations.
They show a gap $\Delta_{cSDW}$ (\ref{delta_csdw}) at cSDW momenta.
Those originating from the second term in (\ref{G})
are combinations of a hole with a spinwave,
with a total energy that lies below the Fermi level.
The incoherent contribution originating from the first term in (\ref{G})
is the time-reversed counterpart, 
and it doesn't contribute 
at energies below the Fermi level in the zero-temperature limit.
Last, the ratio of the incoherent spectral function integrated over momentum 
in the vicinity ${\bf k} = 0$ or
$(\pi/a) {\bf\hat x}$(${\bf\hat y}$)
compared to the coherent counterpart is
$\sum_k^{\prime} {\rm Im}\, G_{\rm inc}(k,\omega) / \sum_k {\rm Im}\, G_{\rm coh}(k,\omega)
= (\gamma / 4\pi s_0) \cdot [(-\omega)\Omega_{\parallel} / (v_0 / a)^2]$
as $\Delta_{cSDW}\rightarrow 0$,
where $-\omega$ measures how far in energy the hole lies below the Fermi level,
and where $\gamma$ is the anisotropy parameter of the spinwave dispersion in question
(see fig. \ref{sw_spctrm}a).

We will now evaluate the former incoherent contribution 
to the spectral function in the large-$s_0$ limit
as  $\Delta_{cSDW}\rightarrow 0$,
at energies just below the Fermi level.
The previous long-wavelength approximations
for the spinwave dispersion near zero 2D momentum
and near cSDW momenta (\ref{sdw_dispersion})
are then valid.  Also valid is the longwavelength approximation for the dispersion of
the slave fermions,
$\varepsilon_f ({\bf k}) = (2 s_0) (-t_1^{\parallel}) (|{\bf k}|^2 - k_F^2)a^2$.
The imaginary part of the pole in the second term of Eq. (\ref{G}) 
enforces energy conservation, 
$-\omega = \omega_b({\bf q}) + \varepsilon_f({\bf q} - {\bf k})$.
The inequality $v_F > v_0$ implies diffuse electron bands centered at momenta
zero and $(\pi/a) {\bf\hat x}$(${\bf\hat y}$)
 that close out at an energy $k_F\cdot v_0 / \gamma$ below the Fermi level.
Figure \ref{photo} shows the net spectral function in these regions at the QCP.
The incoherent contribution was evaluated in
the thermodynamic limit by integrating the $\delta$-function 
in energy over radial momentum analytically,
and by performing the remaining angular integral numerically.
Figure \ref{photo}a shows
that the electronic structure at zero 2D momentum
is predominantly hole-type
because of the coherent contribution.
It is roughly consistent with the
hole Fermi surface pockets about zero 2D momentum
revealed by ARPES 
in the iron-pnictide superconductor BaFe$_{2-x}$Co$_x$As$_2$ \cite{fink09}\cite{brouet11}.
(See fig. \ref{photo}, caption.)
Figures \ref{photo}b and \ref{photo}c show
a mix of electron and hole structure
at cSDW momentum  $(\pi/a) {\bf\hat x}({\bf\hat y})$.
The ``V'' shape that separates pink from purple in fig. \ref{photo}b
and that separates  purple from black in fig. \ref{photo}c defines
an electron Fermi velocity that coincides with the 
cSDW spinwave velocity
along the corresponding principal axis: $v_{l}=v_0$ and $v_{t}=v_0/\gamma_{cSDW}$.
ARPES measurements reported in ref. \cite{brouet11} also reveal 
electron Fermi surface pockets centered at cSDW momenta, 
with Fermi velocities that show the same type of anisotropy.
The electron Fermi velocities extracted from the inner edge of the ``V''
in their dispersion curves,
in particular,
are $v_{l} \sim 0.7$ \AA-eV and $v_{t} \sim 0.3$ \AA-eV.
These values are remarkably close to the corresponding spin-wave velocities extracted
from inelastic neutron scattering measurements on the same material\cite{hayden_10}!
Figures \ref{photo}b and \ref{photo}c also predict stronger hole dispersion 
at 
$\omega = -\varepsilon_f[{\bf k} - (\pi/a) {\bf\hat x}({\bf\hat y})]$, however.
It is due to the divergence of the coherence factor ${\rm sinh}^2 \theta_q$ in Eq. (\ref{G})
as the spin-wave frequency $\omega_b(q)$ vanishes
at ${\bf q}=(\pi/a) {\bf\hat x}({\bf\hat y})$.
Such peaks in ${\rm Im}\, G(k,\omega)$ must therefore disappear
just off the QCP, at cSDW gaps $\Delta_{cSDW} > \Delta k_{peak} \cdot v_0$.
Figures \ref{photo}b and \ref{photo}c yield a peak width for the hole dispersion of
$\Delta k_{peak} a \sim 0.1$.  
The hole-dispersion peaks about cSDW momenta are therefore fragile,
and they may not survive effects such as 
hole (slave fermion) damping and spin-wave (Schwinger boson) damping
that are not accounted for within the present mean-field approximation.

\begin{figure}
\includegraphics[scale=0.70, angle=-90]{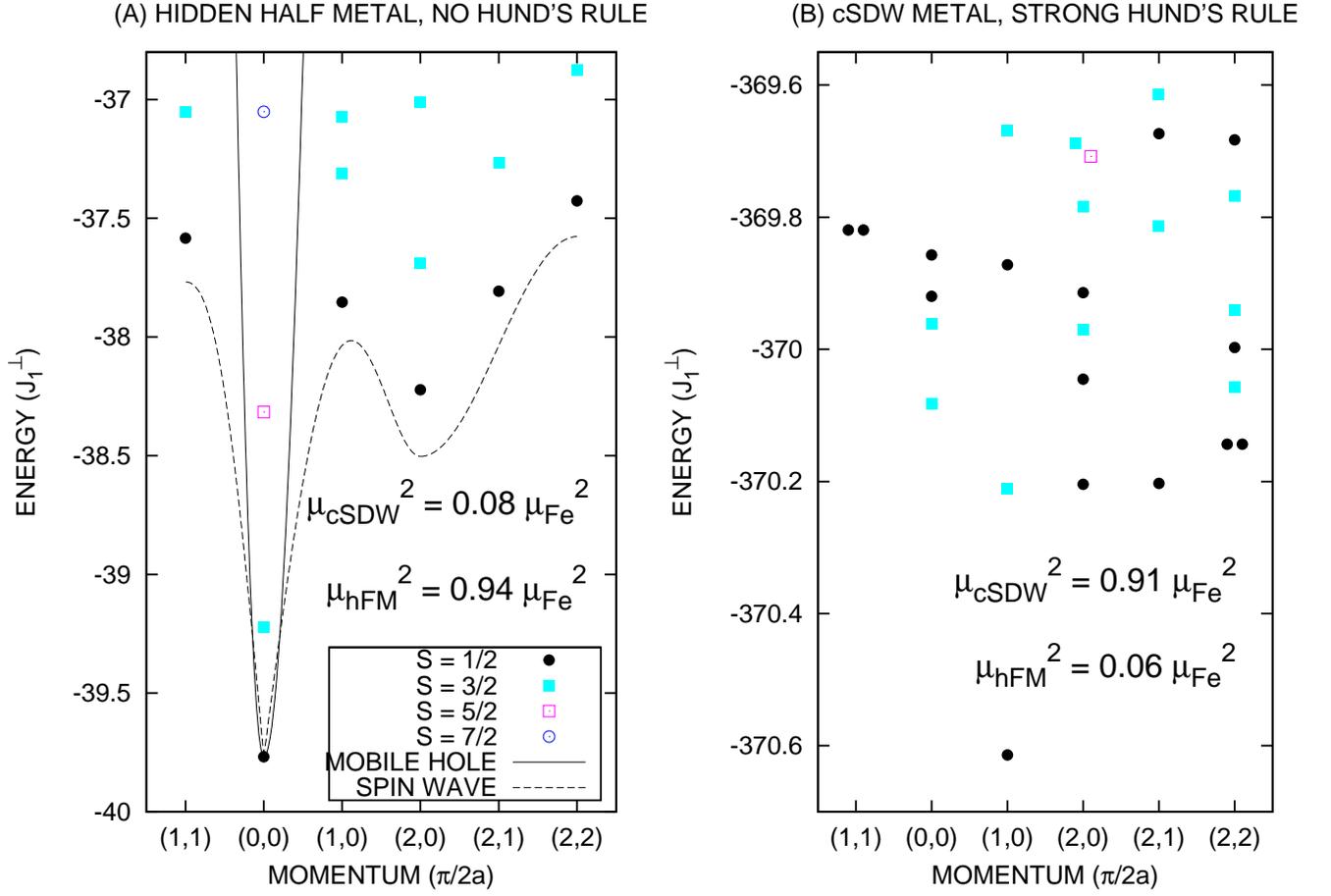}
\caption{
Shown are low-energy spectra  of the $t$-$J$ model,
 Eq. (\ref{tj0j1j2}),  over a
$4\times 4\times 2$  lattice with one hole, with the parameters listed in the text.
Each point is doubly degenerate because $t_{1}^{\perp} = 0$.
A comparison with the hole spectrum, $\varepsilon_f(k)$, and with the spin-wave spectrum,
$\omega_b(k)$, is made (a)
at $x=0$
in the absence of Hund's rule,
$J_0 = 0$.
The latter is enforced (b) by setting $J_0 = - 23\, J_1^{\perp}$.}
\label{exact_spctrm}
\end{figure}

We have confirmed the main results of the above mean field theory
analysis of the hidden half-metal phase
by obtaining the exact low-energy spectrum of one hole in the
two-orbital $t$-$J$ model (\ref{tj0j1j2}) over a periodic  $4\times 4$ lattice of iron atoms
using the Lanczos technique\cite{lanczos}.
The Hilbert space was confined to the $S_z = 1/2$ subspace,
and its dimension was reduced further
by exploiting
translation and reflection symmetries.  
The Heisenberg exchange terms of the Hamiltonian (\ref{tj0j1j2})
were stored in memory as permutations of the spin backgrounds.
Hole hopping was computed at each application of the Hamiltonian (\ref{tj0j1j2}), 
on the other hand.
Last, the Lanczos procedure was applied numerically using
the  ARPACK subroutine library\cite{arpack}.
Figures \ref{exact_spctrm}a, \ref{sw_spctrm}b and \ref{exact_spctrm}b
show how the magnetic order and how
the low-energy spectrum of the $t$-$J$ model  (\ref{tj0j1j2})
evolve with the strength of the Hund's coupling.
Figure \ref{exact_spctrm}a displays a quantitative match
between the dispersion of the exact lowest-energy spin-1/2 excitations
and  the prediction
 for spin-wave excitations
about $\nwarrow_{d+}\searrow_{d-}$ order
at $J_0 = 0$.
Next,
fig. \ref{sw_spctrm}b shows the exact spectrum
at the QCP,
where degenerate groundstates exist at momenta zero and $(\pi/a) {\bf\hat x}({\bf\hat y})$.
The lowest-energy  excitations again have spin 1/2,
and their dispersion is qualitatively similar to
the spin-wave prediction displayed by fig. \ref{sw_spctrm}a.
This quantum-critical state shows weak cSDW order and 
moderate $\nwarrow_{d+}\searrow_{d-}$ spin order.
Last, the Hund's coupling  used in
fig. \ref{sw_spctrm}b is the critical value $-J_{0c} = 2.27 J_1^{\perp}$.
It is considerably larger than the corresponding value 
of $-J_{0c} = 1.35 J_1^{\perp}$ obtained in the absence of a hole
by one of the authors\cite{jpr10}.
This 
is consistent with the previous mean field theory result for $J_{0c}$,
and it is likely due to the suppression of quantum fluctuations in the antiferromagnetic state
by intra-orbital hole motion.
These matches  indicate that
the above Schwinger-boson-slave-fermion mean field theory is
 a valid approximation of the 2-orbital $t$-$J$ model, Eq. (\ref{tj0j1j2}),
 in the hidden half-metal phase.
Finally, fig. \ref{exact_spctrm}b demonstrates that the groundstate
is a robust cSDW state with a large ordered moment
if  Hund's rule is obeyed.
It  carries  momentum $\pm (\pi/2 a) {\bf\hat x}({\bf\hat y})$ (cf. ref. \cite{kane89}),
however, and it
therefore 
is unable to account for {\it any} of the Fermi surface pockets
that are observed by ARPES on iron-pnictide materials\cite{fink09}\cite{brouet09}\cite{brouet11}.

\section {III. Conclusions}
\begin{figure}
\includegraphics[scale=0.70, angle=-90]{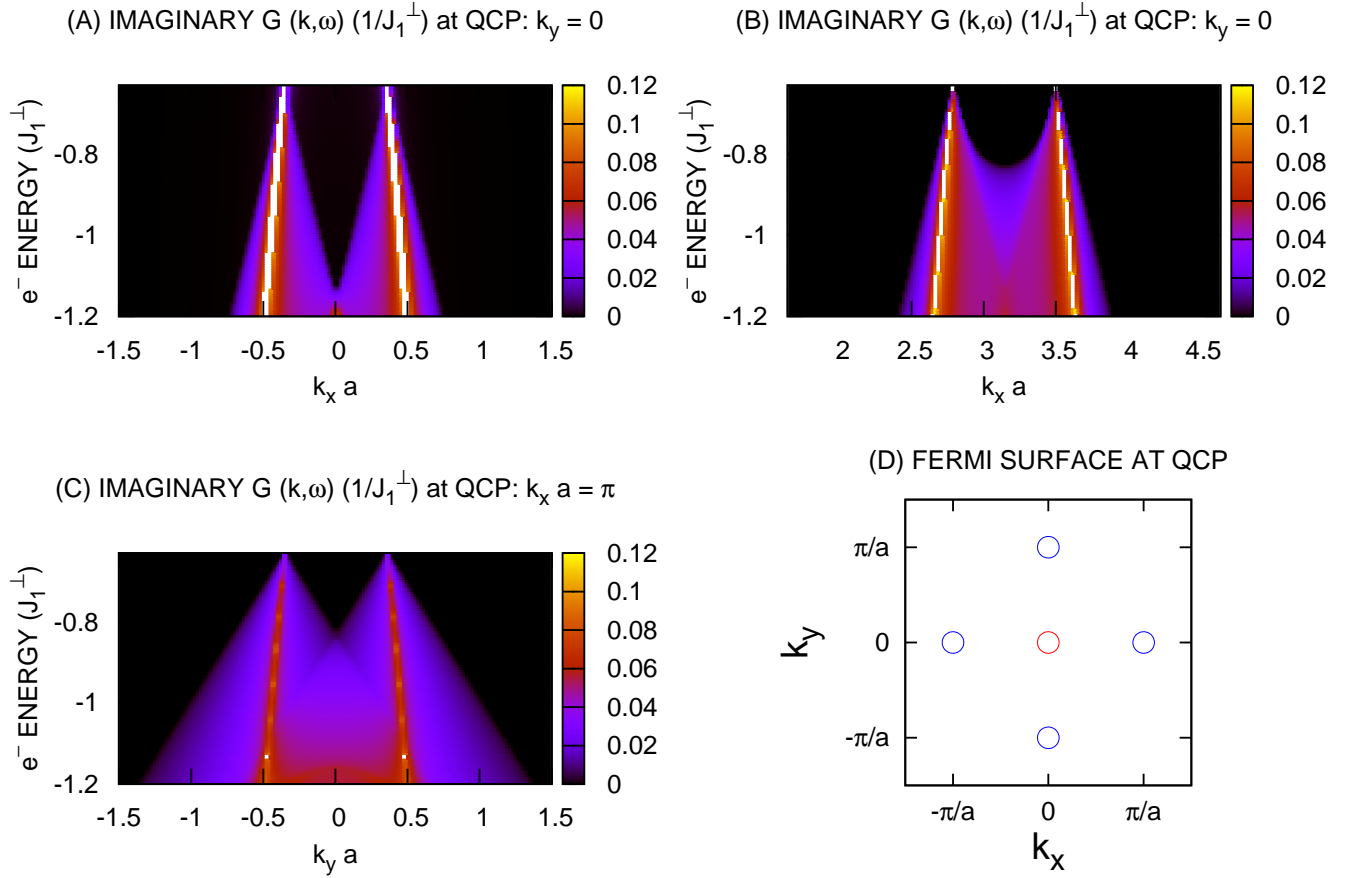}
\caption{Shown is the imaginary part of expression (\ref{G}) 
with the parameter set that is listed in the previous caption,
and in the following limits listed in order:
$T\rightarrow 0$, then $s_0\rightarrow\infty$,
and  then $\Delta_{cSDW}\rightarrow 0$.
The Fe-Fe distance $a = 2.8$ \AA\,
yields the Fermi momentum $k_F = 0.13$ \AA$^{-1}$.
Setting $s_0 J_1^{\perp} \sim  70$ meV (see text and ref. \cite{jpr10})
yields a spin-wave velocity 
$v_0 \sim 0.6$ \AA-eV,
with anisotropy parameter $\gamma_{cSDW} = 2.64$,
and a Fermi velocity
$v_F \sim 1.4$ \AA-eV.
Also shown are the Fermi surfaces predicted by Eq. (\ref{G}).
Each is doubly degenerate because of the $3 d_{xz}$ and $3 d_{yz}$ orbitals.}
\label{photo}
\end{figure}

Above, we have shown that
a hidden half-metal state near a QCP into a cSDW state
exhibits the nested 2D Fermi surfaces that are characteristic of iron-pnictide 
high-$T_c$ superconductors\cite{brouet09}\cite{brouet11}.  
The 
Fermi surfaces predicted here 
with the bare minimum of $3d_{xz}$ and $3d_{yz}$ orbitals
are in fact similar to
those obtained by electronic band structure calculations 
that include all five $3d$ orbitals\cite{graser_09}.
In particular, zone-folded Fermi surfaces centered at 
momentum $(\pi/a , \pi/a)$ have low spectral weight in such case
[cf. fig. \ref{photo}d].
Further, our mean field theory predicts a mixture of electron and of hole dispersion
for Fermi surface pockets centered at cSDW momenta (see fig. \ref{photo}).
This has been been observed by ARPES on iron-pnictide superconductors 
in certain cases\cite{brouet11}.

We have also demonstrated above that the low-energy spectrum of the hidden half-metal state
at the QCP contains zero-energy spin-wave excitations
that disperse anisotropically away from cSDW momenta.  
The   critical spin-wave spectrum notably
exhibits a local maximum at the N\' eel momentum $(\pi/a, \pi/a)$,
which is consistent with INS on a parent compound to  iron-pnictide superconductors\cite{zhao_09}.
It was predicted previously by one of the authors in 
the case where mobile holes are absent\cite{jpr10}.
Last, we have identified the anisotropic spin-wave velocities at cSDW momenta 
with the corresponding electron Fermi velocities around 
Fermi surface pockets centered at cSDW momenta.
A comparison of independent ARPES and INS on the same iron-pnictide compound 
bears out this identification\cite{brouet11}\cite{hayden_10}.

J.P.R. thanks Edward Rezayi, Vitor Vieira, and Maria Jose Calderon for discussions,
and he thanks Veronique Brouet for correspondence.
Exact diagonalization of the $t$-$J$
model (\ref{tj0j1j2}) was carried out on the SGI Altix 4700
at the 
AFRL DoD Supercomputer Resource Center.
This work was supported in part by the US Air Force
Office of Scientific Research under grant no. FA9550-09-1-0660 
and by the FCT under grant PTDC/FIS/101126/2008.

\end{document}